\documentclass[a4paper,
               biblatex,     
               keeplastbox,   
               ]{jacow}
\usepackage{pdfpages,multirow,ragged2e} %
\makeatletter%
	\ifboolexpr{bool{xetex}}
	 {\renewcommand{\Gin@extensions}{.pdf,%
	                    .png,.jpg,.bmp,.pict,.tif,.psd,.mac,.sga,.tga,.gif,%
	                    .eps,.ps,%
	                    }}{}
\makeatother

\ifboolexpr{bool{xetex} or bool{luatex}} 
 {}                                      
 {\usepackage[utf8]{inputenc}}           

\usepackage[USenglish]{babel}

\ifboolexpr{bool{jacowbiblatex}}%
 {%
  \addbibresource{bibliography.bib}
 }{}
\begin{document}

\title{Xsuite: an integrated beam physics simulation framework}

\author{G. Iadarola\thanks{giovanni.iadarola@cern.ch}, R. De Maria, S. {\L}opaciuk,   A. Abramov, X. Buffat, D. Demetriadou,\\
L. Deniau, P. Hermes, P. Kicsiny, P. Kruyt, A. Latina, L. Mether, K. Paraschou, \\
 G. Sterbini, F. Van Der Veken, CERN, Geneva, Switzerland\\
P. Belanger, TRIUMF, Vancouver, Canada\\
P. Niedermayer, GSI, Darmstadt, Germany\\
D. Di Croce, T. Pieloni, L. Van Riesen-Haupt, EPFL, Lausanne, Switzerland
}
	
\maketitle

\begin{abstract}
Xsuite is a newly developed modular simulation package combining in a single flexible and modern framework the capabilities of different tools developed at CERN in the past decades, notably Sixtrack, Sixtracklib, COMBI and PyHEADTAIL. The suite consists of a set of Python modules (Xobjects, Xpart, Xtrack, Xcoll, Xfields, Xdeps) that can be flexibly combined together and with other accelerator-specific and general-purpose python tools to study complex simulation scenarios. The code allows for symplectic modeling of the particle dynamics, combined with the effect of synchrotron radiation, impedances, feedbacks, space charge, electron cloud, beam-beam, beamstrahlung, and electron lenses. For collimation studies, beam-matter interaction is simulated using the K2 scattering model or interfacing Xsuite with the BDSIM/Geant4 library. Tools are available to compute the accelerator optics functions from the tracking model and to generate particle distributions matched to the optics. Different computing platforms are supported, including conventional CPUs, as well as GPUs from different vendors.
\end{abstract}

\section{Introduction}

CERN has a long tradition in the development of software tools for beam physics in circular accelerators. In particular, over the years it has provided the following tools to the user community:
\begin{itemize}[noitemsep,topsep=2pt,parsep=2pt,partopsep=0pt]
\item MAD-X, which has become a standard to describe accelerator lattices, to perform optics calculation, design, and tracking simulations~\cite{madx_ipac23};
\item Sixtrack, a fast tracking program used mainly for long single-particle tracking simulations~\cite{sixtrack_19};
\item Sixtracklib, a C/C++ library for single-particle tracking compatible with Graphics Processing Units (GPUs) accelerators~\cite{sixtracklib_ipac2021};
\item COMBI, a simulation code for the simulation of beam-beam effects using strong-strong modelling~\cite{Pieloni:1259906};
\item PyHEADTAIL, a Python toolkit for the simulation of collective effects (impedance, feedbacks, space charge, and e-cloud)~\cite{pyheadtail_hb16}.
\end{itemize}

These tools were developed over several years, mostly by independent teams. Although they provide very advanced features in their respective domains, with the exception of PyHEADTAIL and SixTracklib \cite{meghana_madhyastha_2019_2551174,Kornilov_2020,PhysRevAccelBeams.24.024201}, their design does not allow effectively combining them for integrated simulations involving complex heterogeneous effects.

Several of these simulations are very well suited for computation acceleration based on GPUs. However, it would be cumbersome to retrofit such a capability in the existing codes.

Furthermore, some of the tools provide their own user interface, consisting in some cases in input/output text files, in some others in an ad-hoc scripting language like in the case of MAD-X. In contrast to this approach, the present de-facto standard in scientific computing is to provide software tools in the form of Python packages that can be easily used in notebooks or integrated within more complex Python codes. This allows leveraging an ever-growing arsenal of general-purpose Python libraries (e.g. for statistics, linear algebra frequency analysis, optimization, data visualization), which is boosted by substantial investments from general industry, especially for applications related to data science and artificial intelligence.

Based on these considerations, in 2021 the Xsuite project~\cite{xsuite_doc} has been launched to bring the know-how built in developing and exploiting the aforementioned codes into a modern Python toolkit for accelerator simulations, which is designed for seamless integration among the different components and for compatibility with different computing platforms, including multicore CPUs and GPUs from different vendors.

Xsuite has by now reached a mature stage of development and has already been adopted as ``production tool'' for several types of simulations across a quite large user community.

In this contribution, we will first describe the overall code structure and development strategy and then illustrate the main features and applications of Xsuite.

\section{Structure, resources, and development strategy}

Xsuite is composed of six modules:
\begin{itemize}[noitemsep,topsep=2pt,parsep=2pt,partopsep=0pt]
\item \textbf{Xtrack}: provides a single-particle tracking engine, featuring thick and thin maps for a variety of accelerator components, together with tools to load and save beam line models, track particles ensembles, characterize the beamline optics;
\item \textbf{Xpart}: provides functions for the generation of particle distributions matched to the beamline optics;
\item \textbf{Xfields}: provides modules for the simulation of collective effects (space charge, beam-beam, electron clouds);
\item \textbf{Xcoll}: provides tools for the simulation of particle-matter interaction in collimators and other beam-intercepting devices (see also~\cite{frederik_hb2023});
\item \textbf{Xdeps}: manages tasks and deferred expressions for modeling and updating of accelerator circuits and other parameter dependencies, provides a multi-objective optimizer (see also~\cite{szymon_hb2023}) and a general purpose text-based tabular data explorer;
\item \textbf{Xobjects}: provides the low-level infrastructure for memory management and multi-platform code compilation and execution (see also~\cite{szymon_hb2023}).
\end{itemize}

All packages are available in the standard Python Package Index (PyPI), they can be installed with a single command (i.e. \verb^pip install xsuite^) and can be easily updated to the latest version at any time.

The code has been designed bearing in mind that, while running on different hardware platforms and covering a large spectrum of phenomena and applications, the software needs to grow in a ``sustainable'' way, being managed and maintained by a small core team integrating contributions from a wider developer community. 
For this purpose, the project employs an “orthogonal” design strategy, ensuring that each module and functional block remains well-isolated and interacts with the others through clearly defined interfaces. This approach has two key advantages: firstly, it enables contributors to modify or expand specific components without requiring comprehensive knowledge of the other parts nor of the underlying software infrastructure. Secondly, it aims to minimize the overall codebase complexity, ensuring that it increases linearly rather than exponentially as new features are added.

Since the early stages, the developers engaged with the user community, encouraging and supporting the users to test and exploit the available features in full-scale simulations studies and integrating their feedback. The tool evolved incrementally, promptly applying corrections and improvements when needed.
Such an approach is based on a fast release cycle, with new versions released typically multiple times per month, while ensuring that there is no disruption  due to version changes on the user's side. This is made possible by an extensive effort in automatic testing, with each version of Xsuite undergoing over a thousand automatic checks (on CPU and GPU) before being rolled out to production.

Throughout the development process, we put a significant effort into building and maintaining a proper code documentation, which is of great importance when developing a tool that is meant to serve a large user community and to cover a broad set of use cases. Xsuite offers different complementary documentation sources~\cite{xsuite_doc}:
\begin{itemize}[noitemsep,topsep=2pt,parsep=2pt,partopsep=0pt]
\item A \textit{User's guide}, describing how to install and use the code for different applications;
\item A \textit{Physics guide} describing the implemented physical models and numerical methods;
\item A \textit{Developer's guide} providing information on the code internals and describing how to add new features;
\item \textit{Docstrings} associated to each class and method, which are accessible directly in the Python interpreter or notebook and are collected in the \textit{Xsuite API reference}.
\end{itemize}

\section{Available features and their applications}

\subsection{Lattice modelling and single particle tracking}

The beam line is represented as a sequence of Python objects from the Xtrack module, each corresponding to an accelerator element or to other physical processes (e.g. magnets, cavities, aperture restrictions, etc.). The model can be defined manually by the user or imported from MAD-X, accounting for element tilts, misalignments, which are handled as changes of the reference system, and multipolar errors.

The implemented models are largely based on the Sixtrack and Sixtracklib implementations, where a `thin` lattice integration method is adopted. Additionally,  `thick` models are available for bending magnets and quadrupoles. In particular, for bending magnets, it is possible to choose between a `full` map, which is appropriate for small accelerators~\cite{Forest_book1}, or an `expanded` one, which is appropriate for large machines and small bending angles~\cite{Ripken:281283}. Dipole edge effects (including fringe fields) can be included either in their linearized form or as full non-linear maps (using the same fringe model as in MAD-NG and PTC~\cite{madng_doc}).

To speed up the simulation, Xsuite assembles and compiles a C kernel callable from Python, which is able to track the entire beamline on CPU or GPU. The tracking speed is found to be similar to Sixtrack for single-core CPU and about two orders of magnitudes faster than that on high-end GPUs~\cite{Hermes_ipac22}. Developments are ongoing to deploy Xsuite on the LHC@Home volunteer computing platform~\cite{lhcathome_web}.

Xsuite tracking simulations have been already used for several applications, notably slow extraction studies~\cite{pniedermayer_ipac23, medaustron_ipac23, sis18_ipac23, ps_slow_extraction_ipac23}, Dynamic Aperture and lifetime simulations ~\cite{pugnat_hb2023}, simulations of halo cleaning using hollow electron lenses~\cite{Hermes_ipac22, rakic_ipac23}, studies on chaos indicators for non-linear beam dynamics~\cite{Montanari_ipac2022}, dynamic aperture estimates based on machine learning~\cite{VanderVeken_ipac22}. 

\subsection{Dynamic control of beam line parameters}

\begin{figure}[t]
   \centering
   \vspace*{0cm}
   \includegraphics[width=.45\textwidth, trim={0.1cm .8cm 1.3cm 0.5cm}, clip]{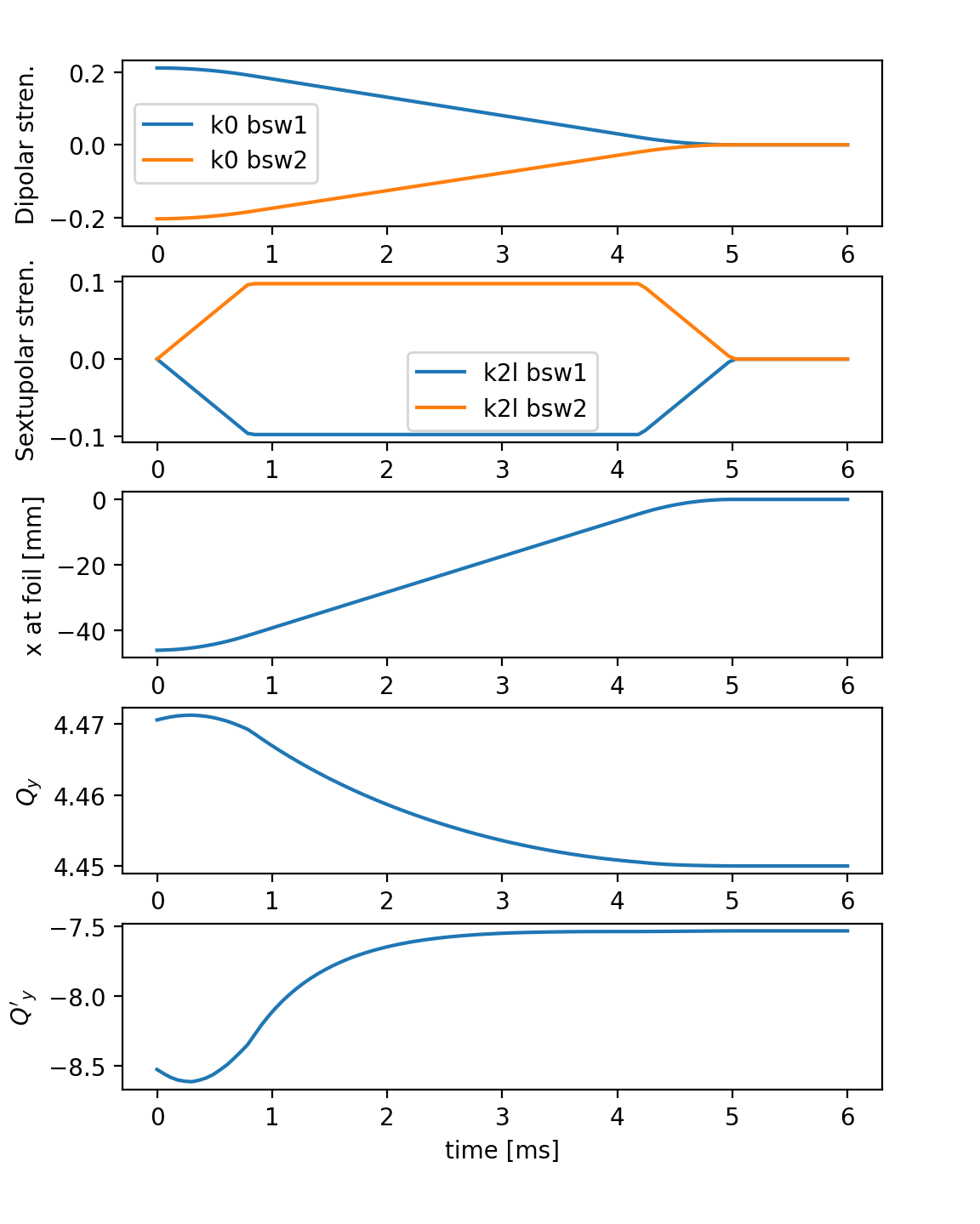}
   \caption{\small Simulation of a fast orbit bump used for the H$^-$ injection into the CERN PS Booster. Form top to bottom: 1.~Time dependence of the dipolar strength of two of the dipole correctors; 2.~Corresponding sextupolar components from induced eddy currents; 3.~Resulting position at the stripping foil as computed by Xsuite; 4.~Effect of the dipole change on the vertical tune (mostly due to fringe fields); 5.~Effect of the dipole change on the vertical chromaticity. }
       \vspace*{-.5cm}
   \label{fig:psb}
\end{figure}

Accelerators and beam lines have complex control patterns. For example, a single high-level parameter can be used to control groups of accelerator components (e.g. sets of magnets in series, groups of RF cavities, etc.) following complex dependency relations. The Xdeps module provides the capability to include these dependencies in the simulation model so that changes in the high-level parameters are automatically propagated down to the line elements properties (see also~\cite{szymon_hb2023}).

For example, as in MAD-X, the user can set the crossing angle between the two colliding beams for the LHC by setting 
``\verb|lhc.vars['on_x1'] = 160 # murad|'',
which automatically changes the strength of 40 dipole corrector magnets in the two beam lines to control the angle of the two beams at the interaction point and compensate the resulting spurious dispersion.
 Dependencies among parameters can be defined directly by the user or imported from the MAD-X model and can be easily inspected and modified at any time.
Furthermore, it is possible to define "time functions", i.e. time dependent knobs that are updated automatically during the simulation. For example, Fig.\,\ref{fig:psb} shows the evolution of the tune and chromaticity as a result of fringe fields and eddy currents in the fast dipole magnets used for the H$^-$ injection in the CERN PSB.

Dedicated features have been developed to efficiently model noise or ripples in a large number of machine elements as well as fast beam excitations. These have been used, for example, to study different methods for slow extraction and to study the effect of amplitude and phase noise on the HL-LHC crab cavities~\cite{pniedermayer_ipac23, medaustron_ipac23, fornara_ipac23}.

\subsection{Twiss module}

The user can easily obtain the lattice functions of a ring or a beamline using the Twiss method associated to the Xsuite beam line object. The calculation, which probes the lattice simply by tracking suitable particles, is performed through the following steps:
\begin{enumerate}[noitemsep,topsep=2pt,parsep=2pt,partopsep=0pt]
\item The closed orbit is found by applying a standard Python root finder to identify the fixed point of the one-turn map (by tracking a particle at each iteration);
\item The Jacobian matrix of the one-turn map is computed by tracking particles to evaluate the derivatives, using a central-difference formula; \item Lattice functions are obtained by computing the ``Linear Normal Form'' of the map from the eigenvalues and eigenvectors of the Jacobian matrix~\cite{Wolski_book}; 
\item Particles tracking is used to propagate the eigenvectors along the beam line;
\item Twiss parameters ($\alpha$, $\beta$, $\gamma$), dispersion functions, phase advances, as well as the effect of linear coupling are obtained using the 
Mais-Ripken approach~\cite{Willeke:194174}.
\end{enumerate}

The accuracy of such a method is found to be excellent for all accelerators tested so far. For example, for the LHC, the computed beta functions are consistent with the values computed by MAD-X up to the fifth significant digit.
The Twiss computation time is similar to other tools used for the same purpose. For example, for the LHC, the Twiss computation takes a similar time compared to MAD-X.

The Twiss computation can be optionally performed on a portion of the beam line with given initial conditions, and for an assigned beam momentum, which allows studying off-momentum beta-beating, non-linear chromaticity etc.

The fact that the computation of the Twiss parameters is based on the tracking engine itself provides two main advantages. Firstly, the effect of any physical model included in the tracking is automatically accounted for in the Twiss calculation without additional code development effort and including cases where no analytic expression is available for the map, e.g. for interpolators. Secondly, this makes the Twiss a powerful diagnostics tool on the built tracking model, allowing to measure key quantities like tunes, chromaticities and closed orbit on the tracking model itself directly, effortlessly and without exporting or manipulating the model. This proved to be an invaluable tool for identifying and investigating issues with user's models and in the code itself. 

The results of the Twiss method are also used to generate particles distributions matched to the optics, and to configure optics-dependent parameters in the simulations, e.g. collimator gaps, beam sizes in space-charge and beam-beam elements or grid sizes in electron cloud simulations. 

\subsection{Optimizer}

Accelerator design and simulations often require solving optimization problems. Typical examples consist in setting given quadrupole and sextupole circuits to obtain desired values of the tunes or the chromaticity, or in using multiple orbit correctors to obtain a closed orbit bump at a given location, or in optimizing quadrupole strengths to fulfill given constraints on the lattice functions (optics matching).

For this purpose, Xsuite provides an optimizer module to "match" model parameters to assigned constraints, which is built based on the extensive experience of MAD-X. The chosen optimization algorithm is the same implemented in MAD-X~\cite{jacobian_de_maria_icap06}, which has proven over the years to be very well suited for accelerator design and tuning.

The optimizer interface is designed to maximize usage flexibility. In particular, the user can interact with the program during the optimization process, by enabling/disabling targets or knobs, changing target values and tolerances, rolling back optimization steps, and changing knob limits. It is possible to perform optimizations involving targets and knobs from multiple beam lines. The optimizer can also be used to synthesize knobs controlling multiple parameters, as in the case of the crossing angle knob illustrated above.

\begin{figure}[t]
   \centering
   \vspace*{0cm}
   \includegraphics[width=.5\textwidth, trim={0.cm .2cm 0cm 0.5cm}, clip]{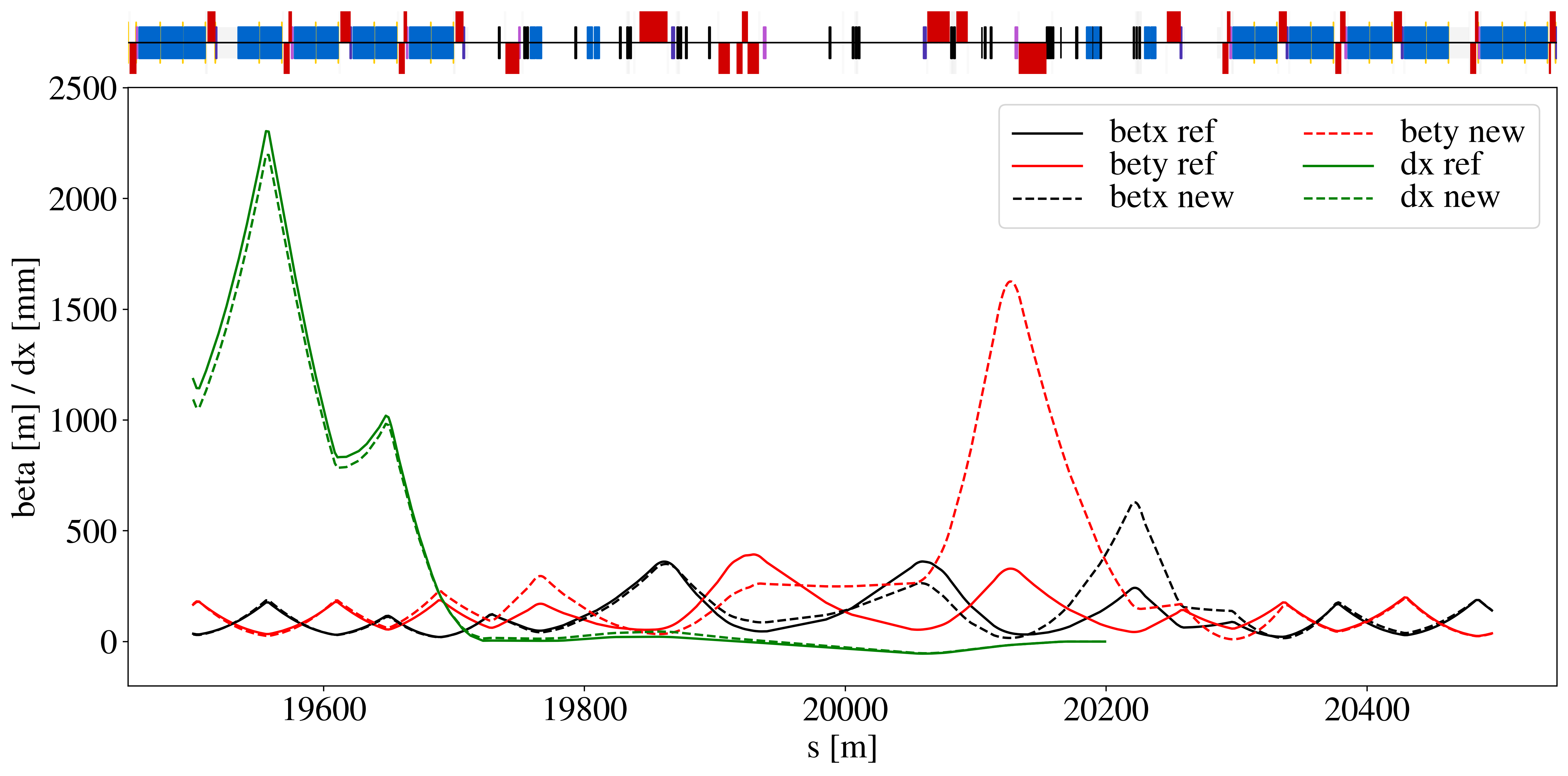}
   \caption{\small Novel optics for the LHC betatron collimation insertion (IR7) designed using the Xsuite optimizer to improve the beam cleaning efficiency and decrease the contribution of the collimators to the transverse impedance of the ring (courtesy of B. Lindstrom~\cite{bjorn_hb2023}).}
       \vspace*{-.5cm}
   \label{fig:ir7optics}
\end{figure}

While by default the targets consist of selected outputs of the Twiss calculations, it is possible to define custom targets and actions involving arbitrarily complex operations. For example, a special ``Luminosity'' target is defined to control beam separations in a collider to obtain a desired luminosity.
In another application shown in Fig.\,\ref{fig:ir7optics}, a custom action tracking particles within the LHC collimation area, has been used to optimize the optics functions in order to maximize the efficiency of the collimation system~\cite{bjorn_hb2023}.
The Xsuite optimizer has also been used for optics matching of the LHC and of FCC-ee collider, showing its capability of handling large problems with several constraints and degrees of freedom.

\subsection{Synchrotron radiation models and compensation}

The effect of synchrotron radiation can be included in Xsuite tracking simulations. For this purpose, the user can choose between two models:
\begin{itemize}[noitemsep,topsep=2pt,parsep=2pt,partopsep=0pt]
\item The ``mean'' model, for which the energy loss from the radiation is applied particle by particle without accounting for quantum fluctuations;
\item The ``quantum'' model for which the actual photon emission is simulated including quantum fluctuations, using the algorithm described in~\cite{helmut_synrad_90}.
\end{itemize} 
Presently, these features are implemented only for the ``thin'' lattice modelling.

\begin{figure}[t]
   \centering
   \vspace*{0cm}
   \includegraphics[width=.44\textwidth, trim={0.cm .2cm .5cm 0.5cm}, clip]{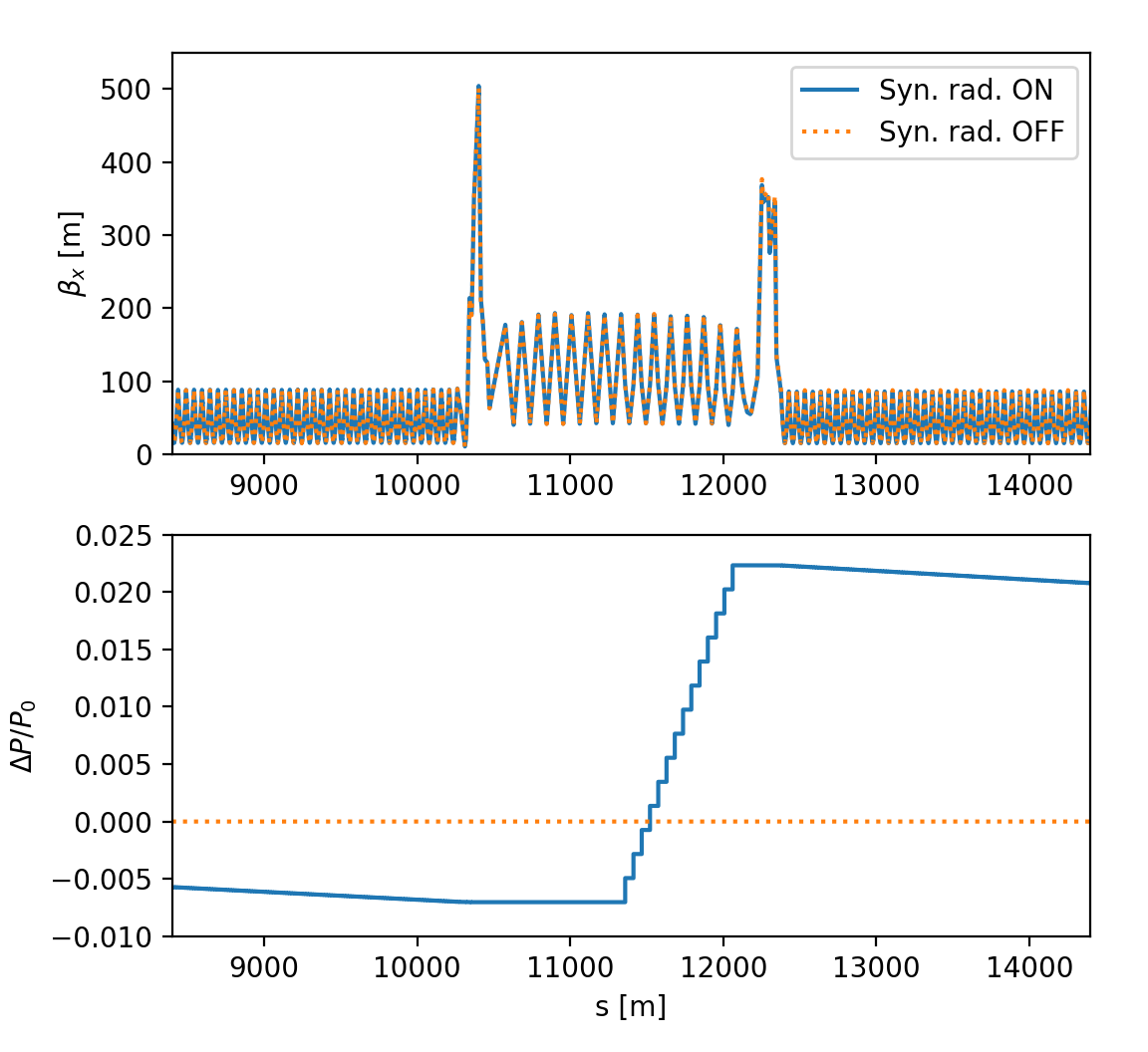}
   \caption{\small Optics and beam energy deviation in the area close to one of the RF insertions of the FCC-ee collider plotted in the absence of radiation (orange) and after the energy loss compensation (blue).}
       \vspace*{-.5cm}
   \label{fig:tapering}
\end{figure}

When synchrotron radiation is present, additional calculations can be enabled in the Twiss calculation. In particular, the energy loss from synchrotron radiation is measured along the beam line, the damping times for the longitudinal and transverse planes are computed from the eigenvalues of the one-turn matrix, and the equilibrium emittances are calculated from the lattice linear normal form following the approach described in~\cite{chao_eq_emit}.

In the presence of radiation, the one-turn matrix is not symplectic, hence the lattice functions cannot be calculated using the conventional Mais-Ripken approach. Instead, two alternative methods are provided by Xsuite to compute the lattice functions in the presence of radiation:
\begin{itemize}[noitemsep,topsep=2pt,parsep=2pt,partopsep=0pt]
\item the ``full'' method, which computes the De Moivre representation of the one-turn matrix from which the lattice functions can be extracted~\cite{Forest_book2};
\item the ``kick as closed orbit'' method, which uses modified tracking maps that measure the momentum changes from synchrotron radiation on the closed orbit and apply the same kicks to all particles. This provides a symplectic one turn matrix accounting to first order for the energy loss from the radiation, on which the Mais-Ripken approach is then applied.
\end{itemize}

The second method is used by default since, for most cases of interest, it provides an excellent compromise between accuracy and computation speed.

In high-energy lepton rings, the beam looses significant energy due to synchrotron radiation. RF cavities around the ring need to be correctly phased to compensate for the loss and the strength of magnetic elements needs to be adjusted to match the actual energy of the beam at the magnet location (this operation is often called ``tapering''). These two corrections are interdependent and cannot be done sequentially.
Xsuite provides an automatic iterative method that obtains the compensation by performing the following steps:
\begin{enumerate}[noitemsep,topsep=2pt,parsep=2pt,partopsep=0pt]
\item Find the 4D closed orbit with no radiation and frozen particle energy along the ring;
\item With radiation enabled, track a particle on the closed orbit while adjusting the strength of each magnet to the actual energy of the incoming particle, which allows measuring the energy loss while preserving the optics and the orbit
\item Adjusts phases of RF cavities to compensate the energy loss, distributing the energy loss according to the cavity voltage;
\item Repeat 2-3 until convergence is achieved.
\end{enumerate}

Tracking and Twiss with radiation as well as the energy loss compensation method are  presently in use for studies within the FCC project, handling energy loss per turn as high as 5\% of the total beam energy. Figure\,\ref{fig:tapering} shows the optics and the beam momentum in a region close to one of the RF insertion of the FCC-ee collider. One can observe how the optics is perfectly preserved by the energy compensation algorithm.

\subsection{Particle-matter interaction and collimation studies}

Xsuite provides dedicated features for simulating particle-matter interactions and to study beam collimation.

The tools to model the interaction of beam particles with intercepting devices like collimators, targets, dumps, crystals are provided through the Xcoll package~\cite{frederik_hb2023}. The interaction is simulated using one of the following engines:
\begin{itemize}[noitemsep,topsep=2pt,parsep=2pt,partopsep=0pt]
\item The ``Everest'' engine embedded in Xcoll, which is an evolution of  the K2 model developed for Sixtrack~\cite{despina_xcoll_ipac23};
\item The ``Geant 4'' engine, which exploits an interface between Xsuite and the Geant4 library~\cite{geant4_doc}, built through the BDSIM library~\cite{Abramov:2022usx, NEVAY2020107200}.
\item The ``FLUKA'' engine (presently being finalized and tested), which exploits an interface between Xsuite and the FLUKA Monte Carlo code~\cite{fluka_doc}.
\end{itemize}
Xcoll also provides facilities to automatically install sets of collimators in the Xtrack beam line and to set the positions of their jaws accounting for the local orbit and beam size as calculated through the Xtrack Twiss module.

\begin{figure}[t]
   \centering
   \vspace*{0cm}
   \includegraphics[width=.47\textwidth, trim={7cm 2cm 7cm 3.5cm}, clip]{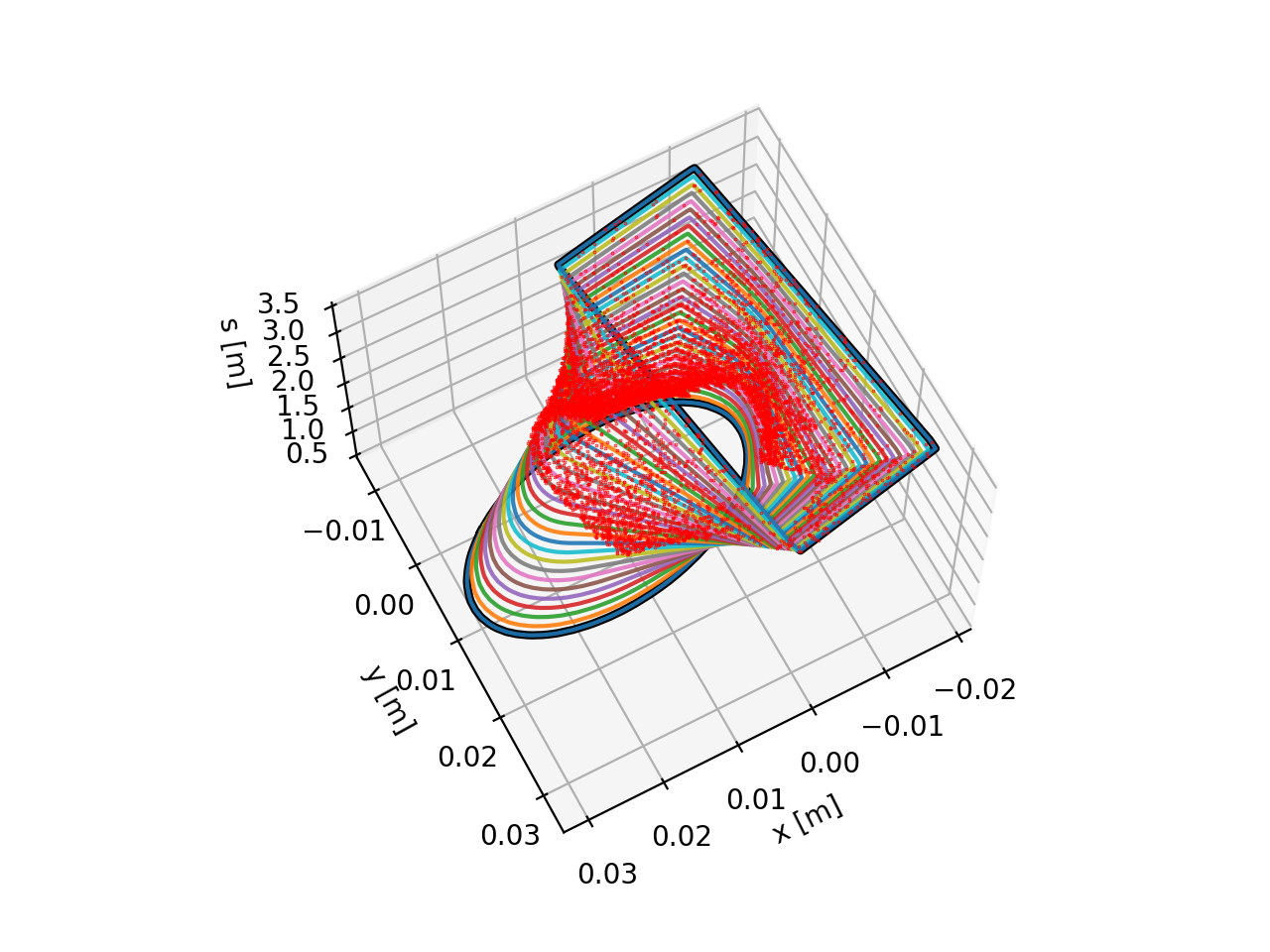}
   \caption{\small Precise localization of the particles lost in the transition between a rectangular aperture and an elliptic aperture, which are shifted and tilted with respect to each other.}
       \vspace*{-.5cm}
   \label{fig:aperture}
\end{figure}

An important goal of simulation studies for collimation is the precise localization of the beam losses along the accelerator, to estimate the power deposition on accelerator components in order, for example, to study equipment activation or quench limits.
The aperture model of the accelerator can be imported by Xsuite as part of the MAD-X sequence. Dedicated elements are installed along the beam line that check the particle positions with respect to the defined aperture and stop the tracking for the particles found to be outside. For collimation studies, such a check is performed at each magnetic element slice, which provides a localization of the losses with a few meters accuracy.
The localization is then further improved in a post-processing stage to reach the accuracy set by the user (typically 1 to 10\,cm), using particle backtracking and a locally refined aperture model obtained by polygonal interpolation (an example is shown in Fig.\,\ref{fig:aperture}).

\subsection{Integration of collective effects}

Collective elements, i.e. elements for which the action on a particle depends on the coordinates of other particles, can also be part of an Xsuite beam line. For such elements, the tracking of different particles cannot happen asynchronously.

In Xsuite, the handling of collective beam elements is fully automatic. The Xtrack line module identifies the collective elements and splits the sequence at the locations of the collective elements, so that the simulation of the non-collective parts can be done asynchronously to gain speed, while the simulation of the collective effects is performed synchronously.

If requested by the collective element, the particles' data is automatically transferred from GPU to CPU for the collective calculation and then moved back.

The simulation of wakefields and transverse feedback systems can be performed by inserting the corresponding PyHEADTAIL elements into the Xsuite line, after enabling a dedicated compatibility layer, which handles the different naming conventions adopted in the two codes.

The effect of space charge and beam-beam, as well as incoherent effects from electron clouds are simulated through the corresponding features of the Xfields module, which will be discussed briefly in the following sections.

\subsection{Space charge}

\begin{figure}[t]
   \centering
   \vspace*{0cm}
   \includegraphics[width=.45\textwidth, trim={0.cm .2cm 0.cm 0.5cm}, clip]{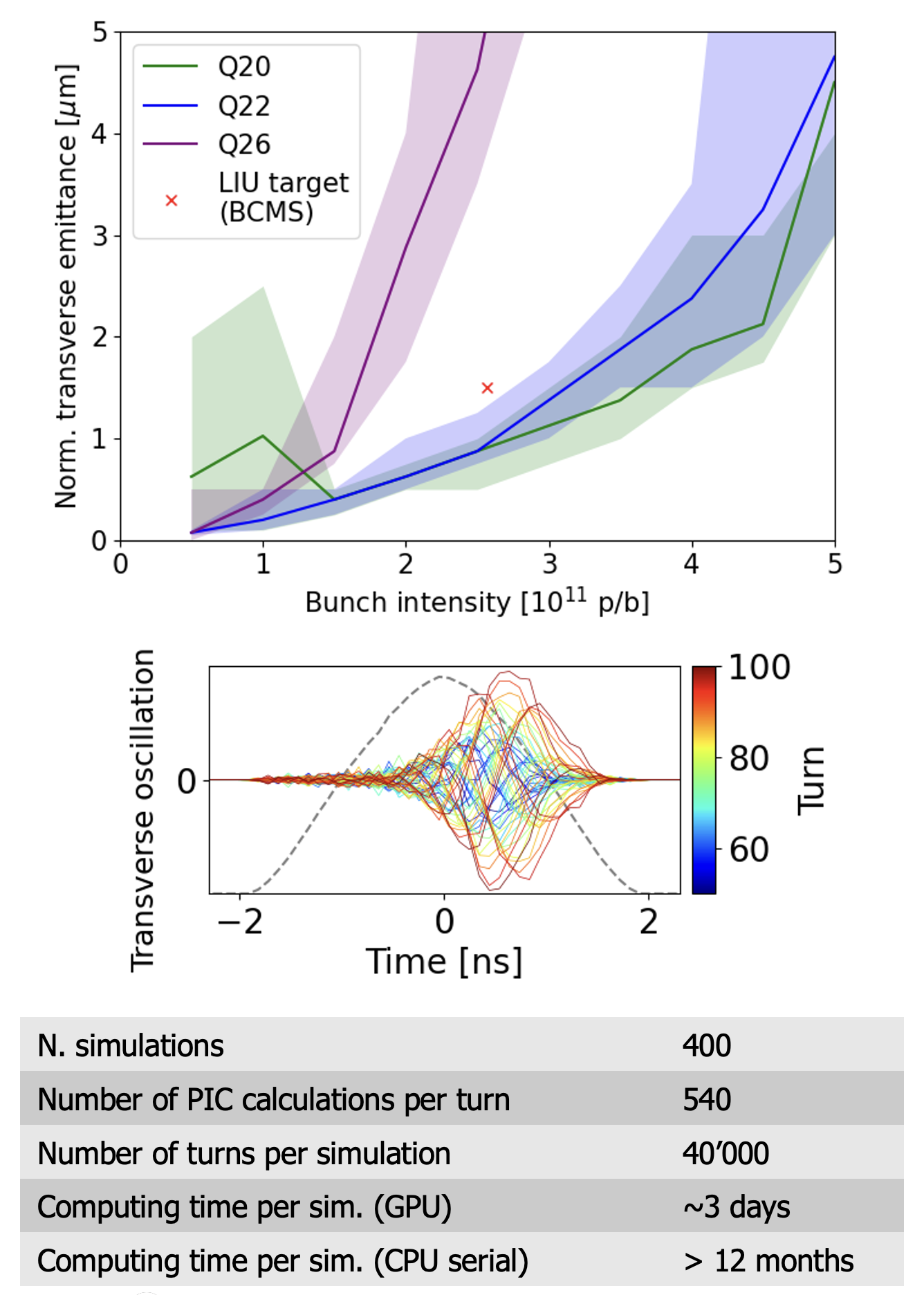}
   \caption{\small Simulation study for the CERN SPS based on simulations including the full non-linear lattice, space charge (PIC model) and wakefields (courtesy of X. Buffat). Top: obtained instability thresholds for different beam optics; middle: one of the simulated instabilities; bottom: some relevant simulation parameters.}
       \vspace*{-.5cm}
   \label{fig:sps_instab}
\end{figure}
The effect of beam space charge can be included in the simulation by inserting dedicated ``lumped'' space-charge elements into the beam line, which apply the space-charge forces to the circulating beam particles. To correctly model the impact of space charge on the beam dynamics, it is necessary to install enough elements to achieve a good sampling of the beam envelope, which typically results in hundreds of space-charge elements installed in the ring. 

Xsuite provides an automatic tool to install and configure the space-charge elements. The user can choose among different space-charge models:
\begin{itemize}[noitemsep,topsep=2pt,parsep=2pt,partopsep=0pt]
\item The ``frozen'' model, in which particles interact with frozen charge distributions that are configured based on the parameters provided by the user (intensity, emittances) and are kept constant over the simulation. In this case, the Xtrack Twiss module is used to compute the optics and obtain transverse beam sizes. The transverse distribution is assumed to be Gaussian.
\item The ``quasi frozen'' model that is a variant of the frozen model in which the beam intensity and beam sizes are recomputed at each interaction from the circulating beam particles.
\item The ``Particle In Cell (PIC)'' model, which consists in the self-consistent computation of the beam field at each space-charge interaction, obtained by distributing the charge from the tracked particle on a uniform rectangular grid and by solving the Poisson equation using a fast solver based on the FFT method with Integrated Green Functions~\cite{QIANG2004278}. The implementation is largely based on the one from PyHEADTAIL-PyPIC\,\cite{pypic_adrian}.
\end{itemize}
Space-charge simulations, especially those employing the PIC method, can be significantly time-consuming. The use of GPUs provides a substantial speed-up.
Figure~\ref{fig:sps_instab} shows the outcome of a study on beam stability for the CERN SPS based on simulations including the full non-linear lattice, space charge (PIC model) and wakefields. In this case, the GPU-accelerated simulations turned out to be more than 100~times faster compared to the serial CPU implementation.

\subsection{Beam-beam interactions}


Similarly to the Sixtrack and COMBI simulation codes, Xsuite provides two models for the simulation of beam-beam effects in colliders:
\begin{itemize}[noitemsep,topsep=2pt,parsep=2pt,partopsep=0pt]
\item The ``4D'' model, which applies only transverse forces independent on the particles longitudinal coordinates, assuming a transverse Gaussian beam distributions;
\item The ``6D'' model, which applies longitudinal and transverse forces accounting for the particles longitudinal coordinates, using the approach described in~\cite{hirata_crossing_angle}, assuming a Gaussian distribution in the transverse plane, and allowing for arbitrary longitudinal beam profile.
\end{itemize}
Both models can be used either in ``weak-strong`` mode, simulating the interaction of the beam particles with a fixed charge distribution modelling the other beam, or in ``strong-strong'' mode where all bunches in the two beams are actually simulated by tracking particles and their interaction is computed accounting for the evolving moments of their charge distribution (updated with a frequency defined by the user).

A tool is provided to automatically identify the locations of the beam-beam encounters, install and configure the beam-beam lenses, based on the geometry of the two rings and on the closed orbit and optics of the two beams as obtained from the Xtrack Twiss. The simulation of beam-beam compensation with wires is also implemented~\cite{sterbini_wire_ipac23}.

In the strong-strong mode, simulations can be very heavy, especially when a large number of bunches needs to be simulated (in the LHC case, thousands of bunches per beam are required). For these cases, the use of parallel computing on HPC clusters is exploited by simulating different bunches on different computing nodes, which exchange information on the distribution of the respective bunches using MPI communication based on the pipeline algorithm~\cite{FURUSETH2019180}.

For the simulation of lepton colliders, the 6D beam element can also account for beamstrahlung and Bhabha scattering. The models implemented for this purpose have been benchmarked against the GUNEA-PIG code \cite{peter_beamstrahlung, peter_bhabha_ipac23}.

Presently, Xsuite has replaced Sixtrack and COMBI as workhorse for weak-strong and strong-strong beam-beam studies for the LHC and HL-LHC~\cite{kostoglou_ipac23} and is being used for beam-beam studies for the design of the FCC-ee lepton collider~\cite{peter_beambeam_ipac23}.

\subsection{Incoherent effects from electron cloud}

Xsuite has been exploited to study the effect of electron cloud on slow beam degradation (emittance growth, losses).

This is done by applying a high-order interpolation scheme, to the e-cloud potential imported on a discrete grid from a dedicated multipacting simulator. 
Also in this case, the use of GPUs is mandatory in order to simulate the required long time scales (millions of revolutions). 

The interpolation scheme is designed to preserve the symplecticity of the resulting map by ensuring the global continuity of the potential, its first derivatives and selected second-order derivatives. More details on these studies can be found in \cite{paraschouIncoherentElectronCloud2020, 
Paraschou:2023ykk, Paraschou:2023tdy}.

\section{ACKNOWLEDGMENTS}

The authors would like to thank F. Asvesta,  H. Bartosik, D. Demetriadou, J. Dilly, C. Droin, S. Fartoukh, A. Fornara, M. Hofer, M. Giovannozzi, S. Kostoglou, P. Kruyt, B. Lindstrom, C. E. Montanari, Y. Papaphilippou, T. Prebibaj, T. Pugnat, S. Redaelli, G. Rumolo, R. Tomas for their important input and support to the development of Xsuite.

\ifboolexpr{bool{jacowbiblatex}}%
	{\printbibliography}%
	{%
	

} 
%
%
\end{document}